\documentclass[conference,compsoc]{IEEEtran}
\IEEEoverridecommandlockouts
\usepackage{cite}
\usepackage{amsmath,amssymb,amsfonts}
\usepackage{algorithmic}
\usepackage{graphicx}
\usepackage{textcomp}
\usepackage{xcolor}
\usepackage{multirow}
\usepackage{booktabs, tabularx} 
\usepackage{subcaption}
\usepackage{float}
\usepackage{centernot}
\def\BibTeX{{\rm B\kern-.05em{\sc i\kern-.025em b}\kern-.08em
    T\kern-.1667em\lower.7ex\hbox{E}\kern-.125emX}}

\begin{document}
\pagestyle{plain}
\title{Impact of Gender on the Evaluation of Security Decisions\\
 }

\author{\IEEEauthorblockN{ Winnie Mbaka}
\IEEEauthorblockA{\textit{Department of Computer Science} \\
\textit{Vrije Universiteit Amsterdam}\\
Amsterdam, Netherlands \\
w.mbaka@vu.nl}
\and
\IEEEauthorblockN{ Katja Tuma}
\IEEEauthorblockA{\textit{Department of Computer Science} \\
\textit{Vrije Universiteit Amsterdam}\\
Amsterdam, Netherlands\\
k.tuma@vu.nl}
}

\maketitle

\begin{abstract}
Security decisions are made by human analysts under uncertain conditions which leaves room for bias judgement. However, little is known about how demographics like gender and education impact these judgments. We conducted an empirical study to investigate their influence on security decision evaluations, addressing this knowledge gap.


\end{abstract}


\section{Introduction}
\label{sec:introduction}

Security decision-makers may encounter cyber challenges that are unfamiliar to them. Yet they are responsible for finding appropriate mitigations while taking into consideration the acceptable levels of risk. The selection of security mitigations and the quality of security evaluations, rely on human judgment and expertise. However, since such decisions  are made in face of uncertainty, the possibility of subjective and biased judgement is introduced~\cite{jaspersen2015probability}.

Indeed judgement bias in risk decision-making has been well documented. For example, evidence from risk analysis literature~\cite{gustafsod1998gender,giddens2020gender} suggests that some demographic parameters (e.g., gender) affect how people perceive security events. In addition, Wright et al. ~\cite{wright2002empirical} observed that one category of experts (e.g., junior vs. senior) might underestimate the feasibility of implementing a particular security countermeasure. Such an event might lead to optimism bias  and could result in not realizing the planned security measure or implementing a less secure workaround. 

Johnson and colleagues \cite{johnson2021decision} present a list of decision-making biases that could be present in cyber activities.
But, the list of biases presented here have not actually been empirically investigated yet.
Measuring judgement bias and human factors in \textit{security decision-making} has not yet been systematically investigated. 

Therefore, our first research question investigates whether the gender and seniority of a security analyst proposing a security mitigation impacts participants' evaluation of the security case study with ethical implications.

\textbf{RQ1:} \textit{Does the perceived gender or seniority of the presented security analyst impact the participant's evaluation of a security case study?}

In addition, the study of bias in isolation fails to provide a more comprehensive perspective of the state of the art in this field. It is important to consider other relevant factors that may have an influencing effect on security decisions, such as ethical concerns or gender stereotypes. It is only recently that studies have begun to investigate the existence of gender stereotypes in computer security and privacy, as in the case of Wei et al. \cite{wei2023skilled}.

To this end, our second research question investigates whether participant's gender or level of education impact their evaluation of the security case study with ethical implications.

\textbf{RQ2:} \textit{Does the gender or education level of the participant impact their evaluation of a security case study?}





\begin{figure}[h]
    \centering
    \includegraphics[width=1\linewidth]{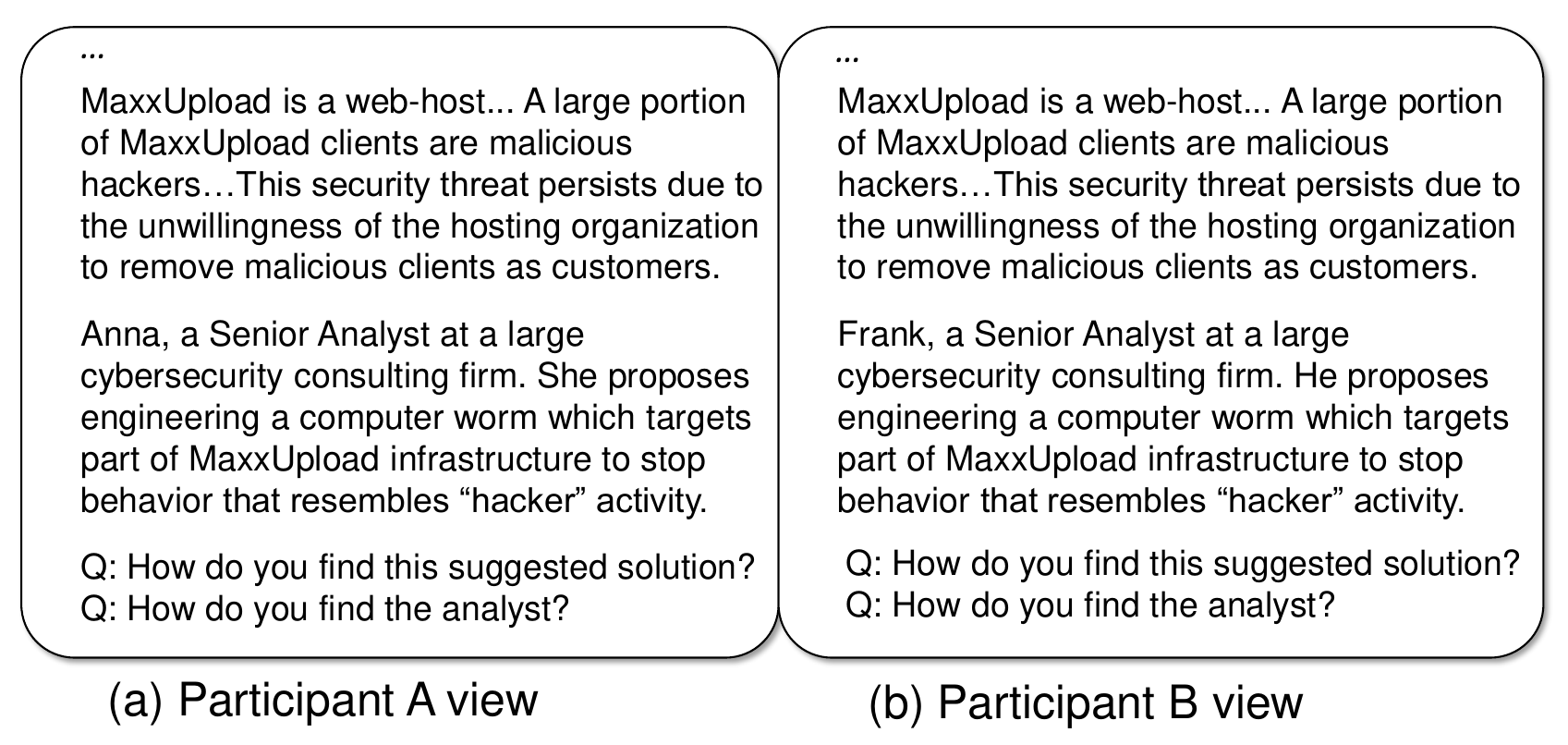}
 
    \caption{Excerpt of the security case study presented to two random participants (different text is emphasized)}
    \label{fig:excerpt}
\end{figure}

\textbf{Our approach.}
Following related work of Wei et al. \cite{wei2023skilled}, as a starting point,  we investigate the impact of binary gender on the evaluation of security decisions.
We conducted a randomised online survey with 188 Bachelor and Master students in computer science.
Figure \ref{fig:excerpt} shows an excerpt of the security case study presented to the participants.
All participants  received the same 
security 
case study, however the description of the vignettes varied slightly. 
Namely, the proposed mitigation by the analysts remained perpetually identical, but the analyst names varied to represent either a female (Anna) or male (Frank) analyst (see Section~\ref{sec:execution}).

\textbf{Our contributions.}
Compared to the related work
we make the following contributions:
\begin{itemize}
\item  We present the first randomised survey investigating bias in evaluating a security case study with ethical implications when considering certain human factors (gender, seniority, and level of education).
\item  We contribute with conducting a survey experiment with 188 computer science university students. We leverage the ACM Malware ethics case study \cite{acmscenario2018} and provide a replication package including the experiment material and analysis protocol \cite{replication2023}. 

\end{itemize}
 We found that participants perceived the analyst personas as equally skilled (statistically equivalent) regardless of their gender or level of education.
In addition, we found that female participants perceived the web-host as (significantly) less ethical and the proposed security mitigation as (significantly) less agreeable.

\section{State of the art}
\label{sec:related}
Table \ref{tab:literature} shows of a summary of the existing literature on the impact of gender and other demographic dimensions 
in risk analysis, security awareness and behavior, and cybersecurity decision-making research. 

\textbf{Risk perception.} In the risk analysis literature, different risk perceptions are already well understood. However,
non-technical scenarios are used to elicit the impact of participants' demographics on risk perception.
For instance, the data for the first study in Table \ref{tab:literature} was obtained 
with
a questionnaire 
including
questions 
about participant's
perception of smoking, stress, and genetically modified organisms.

\textbf{Security awareness and behaviour.} This literature aims to investigate different levels of security awareness and explain individual security behavior online. 
However,
the studies investigating gender present contradicting findings regarding its impact on information security awareness (ISA).
For instance, the first study grouped under this category in Table~\ref{tab:literature}
found that male participants reported higher ISA scores, while the second study reported that female participants had higher scores. 

\textbf{Cybersecurity decision-making.} The literature in this category investigates the gamefication of security decision-making process and the impact of decision-makers level of expertise. Similar to the literature on security awareness and behavior, these studies also present contradicting findings.
On the one hand, 
an individual's
security background knowledge was found to impact 
their
decision-making process. 
On the other hand, no significant impact was recorded when comparing the level of security expertise (experienced vs. inexperienced participants).  

The existing literature discussed shows a correlation between human factors and their associated security practices. However, we observe two significant issues, first, the findings in risk perception literature are derived from non-technical viewpoints. Second, the literature concerning the impact of human factors in security decision making is still in its infancy. 
To this end, our study aims to evaluate the impact of both gender and education level when evaluating security decisions 
when presented with a technical scenario.


\begin{table*}
\caption{Existing literature on the impact of human factors in relevant topic domains}
    \centering
    \begin{tabular}{|p{0.12\textwidth} |p{0.14\textwidth} |p{0.25\textwidth} |p{0.4\textwidth}|}
    \toprule
         \textbf{Topic of research} & \textbf{Demographic measured} & \textbf{Existence of impact} & \textbf{Key references}\\
         \toprule
        Risk perception & Gender, 
         nationality & 
        Non-Swedish \begin{math}
            >
        \end{math}
         Swedish
        
        & A. Olofsson and S. Rashid, “The white (male) effect and risk perception: can equality make a difference?” Risk Analysis: An International Journal
        \\
        
         & Gender, cognitive bias & Male security misbehavior 
         $>$ Female security misbehavior.
         & L. Giddens et al.
         “Gender bias and the impact on managerial evaluation of insider security threats,” Computers \& Security
         \\

 
        \midrule
        Security awareness and behaviour & Gender & 
        Male awareness \& knowledge sources $>$ Female awareness \& knowledge sources
        & A. Farooq et al.
        “Observations on genderwise differences among university students in information security awareness,” International Journal of Information Security and Privacy 
\\
         & Age, gender, personality, and risk-taking propensity & Females and older adults ISA $>$ Males and younger adults ISA 
         & A. McCormac et al.,
         “Individual differences and information security awareness,” Computers in Human Behavior
         \\
        & Gender &  
        Female security behavior \& perceived severity of event $>$ Male security behavior \& perceived severity of event
        & T. McGill and N. Thompson, “Gender differences in information security perceptions and behaviour,”  Australasian Conference on Information Systems
 \\
         \midrule
        Cybersecurity decision making & Security background and knowledge  &  
        Greater security knowledge $\centernot\implies$ better decision making
        & S. Frey, et al.
        “The good, the bad and the ugly: a study of security decisions in a cyber-physical systems game,” IEEE Transactions on Software Engineering
        \\
         &  Professional experience  & 
         Experienced $=$ inexperienced, but Experienced improved decision making overtime
         & M. S. Jalali, et al. 
         “Decision-making and biases in cybersecurity capability development: Evidence from a simulation game experiment,” The Journal of Strategic Information Systems
         \\


         \bottomrule
    \end{tabular}
    
    \label{tab:literature}
\end{table*}


\section{Hypothesis}\label{sec:hypothesis}
We state our hypothesis based on the evidence found in the related literature.
Table \ref{tab:hypothesis} in Appendix \ref{sec:Appendix} summarises the null and alternative hypothesis statistically tested in this study.
\paragraph{\textbf{RQ1}} 
Since no evidence of the contrary was found in the related literature, we expect to find evidence of equally unbiased judgement within the computer science population. 
Therefore, there is no reason to believe that computer science students would judge the analysis outcomes differently depending on the gender or seniority of the security expert who performed the analysis.
Accordingly, we hypothesize finding evidence of statistical equivalence and conduct a Two One-Sided T-Tests (TOST) analysis. 

\paragraph{\textbf{RQ2}} Since some related literature 
(see Table \ref{tab:literature}) found that gender (among other demographic variables) impacts participant's risk perception, we expect to find some differences within our population. For instance, we might observe that female participants perceive security events with greater concern compared to men. 
Therefore, we hypothesize to find evidence of a difference and conduct a test of difference (Mann Whitney U (MWU)).

\section{Experiment artefacts}
\label{sec:artefacts}

We adopt a similar experimental design to the one used by Hibshi et al. \cite{hibshi2015assessment}, a randomized factorial survey. 
Factorial experiments measure the correlation between independent variables to a decision or outcome factor. For our study, the independent variables and outcome factors are outlined in Section \ref{sec:execution}, under measures.
This experimental approach has been used extensively in social sciences
to objectively understand the underlying influences in decision-making. 
During a factorial survey, participants are asked to rate their agreement to a specified outcome that corresponds to a fictitious person. 
To accomplish this, researchers define the dimensions and their levels (which when randomized) comprise a vignette.  
The vignette (which consists of a fictional persona and case study) is then described using potentially relevant characteristics. 

\paragraph{\normalfont{\textbf{Security case study}}}
\label{subsec: scenario}
The scenario is based on the ACM ethics Malware Disruption case \cite{acmscenario2018}. The ACM adopted a new Code of Ethics and Professional Conduct (herein, the Code) in 2018~\cite{acmthecode2018}, which consists of the ethical principles to be respected and six case studies that illustrate how the Code can be applied to situations that arise in everyday practice.
Yet, little is known about how these case studies are actually understood or perceived by members of the computing communities.

To mitigate potential bias associated with the perception of the name "Rogue" (in the original case study) as being negative, we made the decision to change it to "MaxxUpload". 
This modification was aimed at ensuring a more neutral and unbiased representation of the case study. We briefly summarize our adaptation of the case study below. 

\textit{Web hosting organization MaxxUpload is a company that provides hosting services for client websites. A large portion of MaxxUpload clients are hackers who upload malicious programs, which are able to send spam emails containing malicious links. The security threat persists due to the unwillingness of the hosting organization to remove malicious clients as customers.}

\paragraph{\normalfont{\textbf{Mitigation}}}
\label{subsec:mitigation}
Given the security threat presented in the case study
we formulated two mitigations. The first one suggested engineering malware that targets a part of MaxxUpload infrastructure and stops behavior that resembles malicious activity. The second one suggested blocking incoming traffic (except the domains whitelisted as verified non-malicious customers) from MaxxUpload servers. We then dubbed the first and second mitigation as \textit{"Malware"} and \textit{"Traffic blocking"}, respectively.

Each proposed mitigation, malware, or traffic blocking, was also accompanied by a brief explanation of the implications that may occur as a result of implementing the mitigation.
The justification for \textit{"Malware"} was that ~ \textit{This is a corrective type of solution. It entails engineering a computer “worm”, otherwise known as a malware computer program that replicates itself in order to spread to other computers and impact their functionality. Though this solution may work, the worm may target MaxxUpload’s legitimate clients as well, leading to unnecessary loss of data.}

The justification for \textit{"Traffic blocking"} was that ~ \textit{This is a reactive type of solution. It entails coordinating the web browsers with blacklists which block any incoming traffic from MaxxUpload servers. The browsers apply the blacklists by default unless a legitimate MaxxUpload client makes an explicit request to whitelist their specific address. A possible outcome is that all MaxxUpload clients are blocked, and legitimate clients may struggle to reach blacklist maintainers to request that they whitelist their services.}

\section{Considerations about gender}
The pertinent query in our study revolves around how to 
treat
the gender of the analyst suggesting the mitigations and equally significant is the question of what 
information
is 
collected
to 
determine
the gender of the participant.

For signaling the gender of the analyst personas in the vignettes
we used gender trigger names (Frank and Anna) with the intention of implying the gender of the analyst. We made the assumption that our participants would also infer the analyst's gender based on these names, which was accurate for the particular country where the experiment was conducted. This assumption 
is also
true in the home country of the first author (Kenya), however, it is important to note that this assumption may not always apply universally. For instance, in Italy, the name "Andrea" is typically associated with males, while in Germany, the same name is commonly associated with females. Therefore, replicating this experiment in a different country may require a slight redesign of the artefacts (e.g., modifying the names of the analyst).


Regarding the question of how to 
determine
the gender of the participant, we first conducted a preliminary study that included various gender options, male, female, non-binary, and other. From the outcomes of this pilot study, we observed that only six participants identified themselves with gender categories beyond the binary spectrum (3 as non-binary, 2 preferring not to disclose, and 1 specifying "other"). Due to the limited data points associated with each non-binary gender option 
in the pilot study,
we deemed it insufficient to draw meaningful conclusions regarding the impact of non-binary gender on security decisions. 
Similar to the study by Wei et al. \cite{wei2023skilled}, we made a deliberate decision to concentrate our investigation by asking the participants about their sex (as assigned at birth), as a starting point.


\section{Experiment Execution}
\label{sec:execution}

This section presents the steps taken in the execution of the study using the Qualtrics survey tool. 

\paragraph{\normalfont{\textbf{Experiment methodology}}}
Our study adopted the $2^k$ factorial design, \textit{k} is the number of factors to be considered. 
The number of vignettes and their associated attributes used in this study are described in Table \ref{tab:attributes}. 
Each fictitious persona or vignette has 2 dimensions: $\$Name$ and $\$Seniority$ (i.e., position at a large security consulting firm).

\begin{table}[h]
     \caption{The vignette dimensions and levels for the survey designed to measure bias in the judgment.}
    \centering
    \footnotesize
    \begin{tabular}{p{0.2\columnwidth}  p{0.3\columnwidth} p{0.3\columnwidth}}
       \textbf{ Vignette} & \textbf{Gender (Name)} & \textbf{Seniority} \\
    \hline
    SrM  &  Male (Frank)  & Senior Analyst  \\ 
SrF  & Female (Anna) & Senior  Analyst \\ 
JrM  & Male (Frank) & Junior  Analyst  \\ 
JrF  & Female (Anna) & Junior Analyst \\ 
    \hline

    \end{tabular}
    \label{tab:attributes}
\end{table}

For each vignette, a short description of the analyst persona was provided. For instance, \textit{"Anna- a junior analyst"} (inferring their gender and seniority), followed by their suggested security outcome (mitigation).

\paragraph{\normalfont{\textbf{General overview}}}
To carry out the experiment, the vignette persona was randomly assigned to each participant joining the survey, similar to the study in~\cite{hibshi2015assessment}. At the start of the experiment, participants were presented with a consent form.
The participants had a choice to not participate in the study, and their were informed about how the data will be anonymously processed for research purpose.
Next, participants were presented with a description of the security case study, described in Section \ref{sec:artefacts}.
Finally, participants were presented with a security analyst proposing possible security mitigation for the security threat presented by the case study. Alongside the suggested mitigation, the participants could also read a short justification of the presented mitigation to the security issue. In addition, we included a technical appendix describing a practical instance of a security threat (which could be read optionally).




\paragraph{\normalfont{\textbf{Participant target groups}}} For this study, the population of interest is university students enrolled in either a bachelor or master computer science program. Their participation in the study was voluntary.


\paragraph{\normalfont{\textbf{Vignette randomisation}}} To prevent the participants from becoming aware of the research questions, we configured the online survey tool (Qualtrics\footnote{https://www.qualtrics.com/}) to assign each participant with a random vignette persona. 
We configured the randomization of the vignettes such that an equal distribution of the vignettes is maintained.

\paragraph{\normalfont{\textbf{Task}}}
The participants were asked to answer a series of questions about their perception of \textit{\textbf{1)}} the security analyst and \textit{\textbf{2)}} the suggested mitigation.  

As a final step, participants were asked a few questions collecting their demographic information and a series of control questions to rate their understanding of the experiment objects (i.e., proposed solutions, assigned vignettes, analyst justification, and the technical appendix). All responses were captured using a 5-point  Likert scale.

\paragraph{\normalfont{\textbf{Measures}}}
To answer our research questions, we considered independent and dependent variables. The independent variables (or the design of the study) consisted of the variables adopted to infer whether or not biased judgment exists when making security decisions. They include the participants' gender and education background and the levels and dimensions of the vignette persona (i.e., gender and seniority). The dependent (observed) variables of this study are defined measures used to capture the perception of the security mitigations and the analyst persona.

\noindent
\textit{\textbf{Perception of analysts.}} 
To assess perception of the analyst, the participants were asked to rate their confidence in the analyst persona. 
In particular, they were asked to rate five
aspects about the persona, 
namely, they were asked how competent, skillful, knowledgeable, moral, and trustworthy they perceived the analyst persona to be.

\noindent
\textit{\textbf{Perception of the proposed security mitigations.}}
To assess perception of the proposed security mitigation, the participants were asked to rate their confidence in the mitigation proposed by the analyst persona.
The measured variables in this case include, how ethical, reliable, responsible, trustworthy, and overall acceptable they perceived the mitigation to be.  

Participants responses to all perception questions in this study were captured using a 5-point Likert scale\footnote{Where point 1 is labeled "strongly disagree", point 3 is labeled "neutral", and point 5 is labeled "strongly agree".}.

\paragraph{\normalfont{\textbf{Data collection}}} We carried out two data collection campaigns using the same survey structure and experimental materials.


\subsection{Statistical tests}
To analyse data, we perform both test of equivalence and difference. For both statistical tests, Mann Whitney U (MWU) is used with a level of significance equal to 0.05 ($\alpha =0.05$). 

The problem for the equivalence test is formulated as follows;

\begin{eqnarray*}
p_{low}  & =  & MWU (\{x - \delta | x \in A\}, B, alt = 'less') \\
p_{up} & = & MWU (B,\{x + \delta | x \in A\}, alt = 'less')
\end{eqnarray*}




Where $A$ and $B$ are the vectors of the dependent variables (perception of security analyst and mitigation) for each observed demographic dimension of the analyst (gender and seniority) and participant (gender and level of education). The $\delta$, represents the range for which we consider the means of both groups to be equivalent. For both the perception of analyst and mitigation outcomes, we considered the delta of equivalence to be 0.8 (\textit{$\delta$= 0.8}). 
The highest resulting p-value (i.e., between $p_{low}$ and $p_{up}$), determines whether or not the means are equivalent. That is;

\begin{center}
$p_{TOST}$ = max($p_{low}$, $p_{up}$)
\end{center} 

\subsection{Ethical considerations}
\label{subsec:ethical}

This empirical investigation was carried out as part of the course's learning objectives, under the guidance of the experimenters. The artefacts of the survey were presented and discussed in class, following its conclusion. 
We followed the guidelines of the ethical approval process at our institution. Since our study is anonymous, does not collect personal data, merge data sets, or introduce harm (or deception) to our participants, we were allowed to proceed with the study, and the ethical board was not required to get involved for the approval. In addition, we included an informed consent at the start of the experiment and all students agreed to provide their anonymous data for research purpose.

\section{Results}
\label{sec:results}

This section presents the results of the study. Table \ref{tab:demographics} shows the proportion of participants gender and education level 
from both data 
collection campaigns.
In total, 188 students (44 female and 144 male) attending a computer science university program joined the experiment. 
About half of the participants were attending the BSc program (99) and half MSc program (89). 


\begin{table}[h]
\centering
\caption{Demographics}
\begin{tabular}{c|cc|cc}
\hline
      
       &\multicolumn{2}{c|}{Gender}  & \multicolumn{2}{c}{Education level}  \\
       \hline
Total & Male & Female & BSc & MSc \\

\hline
188 & 144  &  44  & 99 & 89   \\ 

\bottomrule
\end{tabular}
\label{tab:demographics}
\end{table}

We have summarised the findings of our investigation in Table \ref{tab:summary of findings}. First, for the effect of the analyst gender or seniority on the participants' perception of the security case study, we did not observe any impact. Second, for the effect of the participants' gender on all the measured variables, we only observed an impact in one instance. That is, on their perception of the ethical implications presented by the case scenario. Third, the participants' level of education did not impact their perception of the analyst persona, the proposed mitigations, or the case study. Lastly, for the impact of the type of mitigation received, we only observed an effect on their perception of the appropriateness of the proposed mitigation to resolve the security problem presented in the case study.

\begin{table}[h]
\caption{Summary of findings. We used symbols to denote the existence (\checkmark), absence of an impact (x), and (-) for instances where we did not investigate impact. } \label{tab:summary of findings}
\centering
\resizebox{0.49\textwidth}{!}{
\begin{tabular}{c|ccc}
\toprule
 &\multicolumn{3}{c}{Perception of:}    \\
      
 & Analyst persona & mitigation & case study  \\

\hline
Impact of analyst gender or seniority & x  &  -  &   -  \\  
Impact of participants' gender &  x &  x  &  \checkmark   \\ 
Impact of level of education &  x &  x  &   x  \\ 
Impact of type of mitigation received &  x & \checkmark &  - \\ 
\bottomrule
\end{tabular}}
\end{table}

The following subsections provides detailed analysis of the results summarised in Table \ref{tab:summary of findings}.

\subsection{Aggregated scale} We aggregated the measured variables into a scale. Namely, we grouped all relevant questions evaluating the security analyst and we grouped all relevant questions evaluating the security mitigation.
We 
verified the reliability of the two aggregated scales (for analyst and mitigation evaluation) using Cronbach alpha.
\paragraph{\normalfont{\textbf{Cronbach alpha}}} 
Cronbach alpha~\cite{Cronbach1951cofficient} is a reliability test which is often used to indicate whether a given scale is consistent or not. 


For the scale used to answer the perception of the analyst persona, we aggregated responses to four questions. They include, how competent, skillful, knowledgeable, and trustworthy (in this case, only for the analyst persona) they perceive the analyst to be. The Cronbach alpha for this scale returned a confidence value of 0.73. We consider our obtained alpha value for this scale to be reliable, as it falls within the range of similar research reports, typically considered dependable when exceeding $\alpha$ = 0.70 \cite{taber2018use}.

For the scale used to answer the perception of security outcomes, we aggregated responses to five questions. They include, how ethical, responsible, trustworthy, and moral 
(in this case, we considered how moral the participants perceived the analyst proposing the outcomes),
and the overall acceptability of the security outcome. The Cronbach alpha for this scale returned a confidence value of 0.85. Similar to our first scale, we consider the resulting value of $\alpha = 0.85$ as reliable \cite{taber2018use}.

Table~\ref{tab:analyst} and \ref{tab:outcomes} in Appendix \ref{sec:Appendix} shows all the measurements before the aggregation. We include them for the purpose of transparency and replicability, however, we have conducted the statistical analysis on the aggregated scales.


\subsection{\textbf{RQ1:} \textit{Does the perceived gender or seniority of the presented security analyst impact participant's evaluation of a security case study? }} To answer the first research question, we 
measured
whether the presented analysts' gender or seniority had an impact on the participant's evaluation of the security case study. 
Table \ref{tab:scale_analyst_demo} shows the mean and standard deviation of the measures.
The participant perception of the analyst and the mitigation outcomes was similar, regardless of gender or seniority.

\begin{table}[h]
\caption{Descriptive statistics on the impact of analyst's gender and seniority on the participant's evaluation of analyst persona and mitigation using the two aggregated scales.
}\label{tab:scale_analyst_demo}
\centering
\resizebox{0.49\textwidth}{!}{
\begin{tabular}{lcrrr|rrrr}
\toprule
 & \multicolumn{4}{c|}{Gender}  & \multicolumn{4}{c}{Analysts' seniority} \\
& \multicolumn{2}{c}{Anna} & \multicolumn{2}{c|}{Frank} & \multicolumn{2}{c}{Senior} & \multicolumn{2}{c}{Junior} \\
 & $\mu$ & $\sigma$ & $\mu$ & $\sigma$ & $\mu$ & $\sigma$ & $\mu$ & $\sigma$ \\
\midrule
  Perception of analyst  &  3.0  & 0.7 & 3.0 & 0.7 & 3.0 & 0.7 & 3.0 & 0.6  \\
  Perception of mitigation  &  2.8  & 1.0 & 2.7 & 0.8 & 2.8 & 0.9 & 2.7 & 0.8  \\
    \bottomrule
\end{tabular}}
\end{table}

%

%
Table \ref{tab:t-test_analyst_demo} is a summary of the resulting two one-sided test for equivalence (the significant results were marked with an asterisk (*)). 
The analysts gender and seniority did not have an impact on the perception of the participants. These results 
support our alternative hypothesis of equivalence of judgement bias.

\begin{table}[h]
\caption{Test of equivalence on perception of analyst and security mitigation when controlling for the gender and seniority of the analyst.}\label{tab:t-test_analyst_demo}
\centering
\resizebox{0.5\textwidth}{!}{
\begin{tabular}{lc|rr|rr}
\toprule
& & \multicolumn{2}{c}{Analyst}  & \multicolumn{2}{c}{Mitigation}  \\

 &  $\alpha =0.05$ & $MWU$ & $P$ & $MWU$ & $P$ \\
\midrule
  Gender  & {$Female$ - $\delta$} $<$ {$Male$}   & 1676.0 & \textbf{8.8e-14*} & 2603.5 & \textbf{5.6e-07*}  ($p_1$)\\

  & {$Male$} $<$ {$Female$ + $\delta$} & 1441.0 & \textbf{6.4e-16} & 2071.5 &  \textbf{1.5e-10}($p_2$)\\

\hline
 Analysts' seniority & {$Senior$ - $\delta$} $<$ {$Junior$} & 1667.0 & \textbf{7.5e-14*} & 2389.5 &  \textbf{2.5e-08*} ($p_1$)\\

  & {$Junior$} $<$ {$Senior$ + $\delta$} & 1462.0  & \textbf{1.0e-15}& 2281.0 &  \textbf{4.8e-09}($p_2$)\\
    \bottomrule
\end{tabular}}
\end{table}

\subsection{\textbf{RQ2} \textit{Does the gender or education level of the participant impact their evaluation of a security case study? }}
To answer the second research question we analysed participants' evaluation of \textbf{\textit{1)}} the analyst persona and security mitigation and \textbf{\textit{2)}} their ethical perception of the case study presented to them.  
We 
measured
whether their gender or their level of education had an impact on their perception of the analyst, the security mitigation, and the case study. 
In addition, 
we
analyzed whether the type of mitigation received had an impact on the participants' evaluation of the security case study. 

\paragraph{\normalfont{\textbf{Perception of the security analyst and mitigation}}}

Table \ref{tab:scale_partcicipants_demo} shows the average responses of the two aggregated scales.
On average, the responses regarding the perception of the analyst persona and the security mitigation were similar, regardless of the participant's gender or level of education.

\begin{table}[h]
\caption{Descriptive statistics on the impact of Participants' gender and level of education on their evaluation of analyst persona and mitigation using the aggregated scales for measures.
}\label{tab:scale_partcicipants_demo}
\centering
\resizebox{0.49\textwidth}{!}{
\begin{tabular}{lcrrr|rrrr}
\toprule
 & \multicolumn{4}{c|}{Participant's gender}  & \multicolumn{4}{c}{Level of education} \\
& \multicolumn{2}{c}{Female} & \multicolumn{2}{c|}{Male} & \multicolumn{2}{c}{BSc} & \multicolumn{2}{c}{Msc} \\
 & $\mu$ & $\sigma$ & $\mu$ & $\sigma$ & $\mu$ & $\sigma$ & $\mu$ & $\sigma$ \\
\midrule
  Perception of analyst  &  2.9  & 0.5 & 2.9 & 0.7 & 2.9 & 0.7 & 3.0 & 0.6  \\
  Perception of mitigation  &  2.8  & 0.9 & 2.7 & 0.8 & 2.7 & 0.9 & 2.8 & 0.8  \\
    \bottomrule
\end{tabular}}
\end{table}



We first conducted a statistical analysis of difference (see Table \ref{tab:scale_diff}).
No statistically significant differences were found.
These results do not provide any evidence to support the alternative hypothesis about perception differences.

\begin{table}[h]
\caption{Test of difference on participants' perception based on their gender and level of education.}\label{tab:scale_diff}
\centering
\resizebox{0.49\textwidth}{!}{
\begin{tabular}{lc|rr|rr}
\toprule
& & \multicolumn{2}{c}{Analyst}  & \multicolumn{2}{c}{Mitigations} \\

 &  $\alpha =0.05$ & $MWU$ & $P$ & $MWU$ & $P$ \\
\midrule
Participant  gender  & {$Female$} $,$ {$Male$}  & 3342.0 & 0.58 & 3557.0 & 0.21  \\
Level of education  & {$BSc$} $,$ {$MSc$} & 4396.0 & 0.98 &  4442.5 & 0.92 \\
    \bottomrule
\end{tabular}}
\end{table}

Since we did not find significant difference and as a confirmation for the similarities observed in Table \ref{tab:scale_partcicipants_demo}, we conducted an equivalence analysis and found statistical equivalence. 
Therefore, we do not observe an impact on participants perception when controlling for their gender and level of education.

\paragraph{\normalfont{\textbf{Perception of the case study}}} Since the participants were presented with a security case study with ethical implications, it is also interesting to investigate their perception of MaxxUpload's business model. That is, the willingness of MaxxUpload to host both legitimate and malicious clients.

Table \ref{tab:scenario_diff} is a summary of the perception of the case study when controlling for the participants gender, level of education, and the type of mitigation received. From this analysis we observed a significant difference when controlling for participants' gender (p-value = \textbf{0.03}). 
In particular, female participants did not agree with Maxxupload's decision to host any type of client (legitimate and malicious ones). 
Since we only found significant difference on one variable, our study can only partially support the alternative hypothesis of difference.

 \begin{table}[h]
\caption{Test of difference on participants' perception of the case study based on their gender and level of education.}\label{tab:scenario_diff}
\centering
\resizebox{0.49\textwidth}{!}{
\begin{tabular}{lc|rr}
\toprule
& & \multicolumn{2}{c}{Case study} \\

 &  $\alpha =0.05$ & $MWU$ & $P$  \\
\midrule
 Participants' gender  & {$Female$} $,$ {$Male$}  &  2514.0 & \textbf{0.03*} \\
Level of education  & {$BSc$} $,$ {$MSc$} &  4807.5 & 0.26 \\

Mitigation  & {\textit{Traffic blocking}} $,$ {$Malware$} &  4167.0 & 0.48 \\
    \bottomrule
\end{tabular}}
\end{table}

For the variables that did not return a significant difference, level of education and mitigation, we confirmed that they are statistically equivalent by performing a MWU two one-sided test. 

\paragraph{\normalfont{\textbf{Perception based on the type of mitigation received}}} In addition to analysing the impact of gender and level of education of the participant on their perception of security analysts and proposed mitigations, we also considered the impact of the type of mitigation received. 
We considered the effect of the mitigation on the perception of the participants towards the analyst persona and the mitigation. 
From the analysis presented in Table \ref{tab:mitigation} we observed a similar perception of the analyst person regardless of the mitigation received. However, there was a slight difference in the perception of the proposed security solution. Namely, participants who received the traffic blocking had a neutral perception, while participants with the malware mitigation had a negative perception. 

\begin{table}[h]
\caption{Descriptive statistics on participant perception when controlling for the type of mitigation received}\label{tab:mitigation}
\centering
\begin{tabular}{lccrr}
\toprule
 
\multicolumn{3}{r}{\textbf{Mitigation;}}\\
& \multicolumn{2}{c}{Traffic blocking} & \multicolumn{2}{c}{Malware} \\
 & $\mu$ & $\sigma$ & $\mu$ & $\sigma$ \\
\midrule
  Perception of analyst  & 2.9 & 0.7 & 2.9 & 0.5 \\
  Perception of mitigation  &  3.0 & 0.7 & 2.4 & 0.8\\
    \bottomrule
\end{tabular}
\end{table}

To this end, we tested for existence of a statistical difference with the results summarised in Table \ref{tab:t-test_mitigation}. We observe a significant difference when controlling for the type of mitigation received (p-value = \textbf{7.8e-07}). We concluded that, participants who received the "Traffic blocking" mitigation perceived it as a more appropriate solution to the security threat present in the case study.

\begin{table}[h]
\caption{Test of difference on participants' perception of security analyst and mitigation based on their gender, level of education and type of mitigation received}\label{tab:t-test_mitigation}
\centering
\resizebox{0.49\textwidth}{!}{
\begin{tabular}{lc|rr|rr}
\toprule
& & \multicolumn{2}{c}{Analyst}  & \multicolumn{2}{c}{Mitigation}  \\

 &  $\alpha =0.05$ & $MWU$ & $P$ & $MWU$ & $P$  \\
\midrule
 Participants' gender  & {$Female$} $,$ {$Male$}  & 3342.0 & 0.58 & 3557.0 & 0.21  \\
level of education  & {$BSc$} $,$ {$MSc$} & 4396.0 & 0.98 &  4442.5 & 0.92 \\
Mitigation received & {\textit{Traffic blocking}} $,$ {$Malware$} & 4397.0 & 0.95  & 6256.0 & \textbf{7.8e-07*} \\
    \bottomrule
\end{tabular}}
\end{table}

Since we did not find significant difference for all the variables, we checked for statistically equivalent. The results in Table \ref{tab:TOST_mitigation} in Appendix \ref{sec:Appendix} confirms their statistical equivalence. 

\smallbreak
\noindent

\noindent

\section{Limitations and conclusions}
\label{sec:validity}
There are several limitation that may limit the validity of this study.

    The small number of female participants that joined our study can be considered a limitation to this study. Unfortunately, the issue of having an unbalanced gender population in STEM discipline is well documented~\cite{rodriguez2021perceived}. To remedy this situation, we encouraged female participants to join the study. in addition, the use of student participants as opposed to expert practitioners is often seen as convenience sampling and presents another limitation. Students are considered to have a limited knowledge in implementing industry-level practices. However, studies such as the one by Svahnberg and colleagues \cite{svahnberg2008using} confirm that students can be used in lieu of experts under certain conditions (i.e., when students have a true commitment to the task). In our case, students were well suited to participate in this experiments as understanding concepts of security mitigations was part of the study objectives of the course taught by the experimenters. 

    We also considered the threat of participants not understanding the presented material, such as the case study. Using a series of control questions, we observed that 90 participants agreed that they understood the security case study while an additional 70 indicated that they strongly agree (point 4 and 5 on the likert scale). When asked if they understood the proposed security mitigation, 53 participants strongly agreed while 90 indicated that they agree. Lastly, when asked if they understood the consequences of the proposed security mitigation, 112 participants responded by agreeing that they do (point 4 on the likert scale) while 42 strongly agreed.




    The aggregated scale used to measure the perception of security outcomes included a variable that was originally considered as a measure for the analyst personas (i.e., a question on morality). However, when checking for reliability of the measured variables using Cronbach Alpha, we found that the confidence level was higher when the responses to \textit{"Moral"} was included in the perception of outcomes and not the analyst persona. We concluded that, the participants considered this questions to refer to the moral consideration of the outcomes presented by the analyst and not the persona themselves. To further solidify our rational for including the question on morality in the perception of outcomes, a Pearson's correlation was performed on all the aggregated variables for each scale. We also consider the limitation presented by arbitrarily estimating the value of delta. Similar techniques have been successfully applied in studies within the pharmaceutical industry and by the Food and Drug Administration (FDA) \cite{meyners2012equivalence} when testing the equivalence of two treatment groups.

    


%

We conducted an experiment with 188 Msc and Bsc computer science students 
we \textbf{\textit{1)}} found evidence of equivalence when analyzing the impact of analysts' and participants' demographic dimensions on the perception of the security analyst persona. 
We also \textbf{\textit{2)}} found evidence of a gender impact in the perception of security threat analysis mitigations and when considering the ethical practices with regards to the web-host (MaxxUpload's) business model. 
The next step in our approach involves treating gender as an intersectional concept. This will involve providing the participants with a full spectrum of gender identities and allowing them to self-identify. In addition, measuring the existence of bias when presented with a gender-neutral/non-binary vignette persona is especially interesting as it contributes to the promotion of diversity in cybersecurity and more generally computer science.

\section*{Acknowledgments}
This work was funded by the \textit{Nederlandse Organisatie voor Wetenschappelijk Onderzoek (NWO)} under the HEWSTI Project under grant no. 14261.

\bibliographystyle{IEEEtran}
\bibliography{references}

\clearpage \onecolumn
\appendix
\label{sec:Appendix}

Online appendix to be available on ArXiv.
\section*{Tested hypothesis}
\begin{table*}[h]
     \caption{Hypotheses}
    \centering
    \small
    \begin{tabular}{p{0.05\textwidth} p{0.45\textwidth} p{0.45\textwidth}}
       \textbf{ Hyp} & \textbf{Null hypothesis} & \textbf{Alternative hypothesis} \\
    \hline
  
    $H1_{equiv}$&  No statistically significant equivalence when comparing participants' evaluation of a security case study when controlling for the gender or seniority of the presented security analyst.   & Participants evaluation of a security case study when controlling for the gender or seniority of the presented security analyst is statistically equivalent.  \\ 
    
    $H1_{diff}$&  No statistically significant difference when comparing participants' evaluation of a security case study when controlling for the gender or seniority of the presented security analyst.   & Participants evaluation of a security case study when controlling for the gender or seniority of the presented security analyst is statistically different.  \\ 
  
$H2_{diff}$& No statistically significant difference when comparing participants' evaluation of a security case study when controlling for their gender or level of education.   & Participants evaluation of a security case study when controlling for their gender or level of education is statistically different.  \\

  $H2_{equiv}$&  No statistically significant equivalence when comparing participants' evaluation of a security case study when controlling for their gender or level of education.   & Participants evaluation of a security case study when controlling for their gender or level of education is statistically equivalent.  \\
    \hline
 
    \end{tabular}
    \label{tab:hypothesis}
\end{table*}

\subsection*{Perception of analyst and security outcomes}
\label{subsec:perception}


\begin{table*}[h]
    \caption{Perception of Analyst Based on Gender (Male, Female) and Level of education (BSc, MSc) of the participants, (\textit{SrM- senior male, SrF- senior female, JrM- junior male, JrF- junior female}, 5-point likert scale)}
    \centering
    \footnotesize
    \label{tab:recognized:threats}
    \begin{tabular}{lrrrrrrrr|rrrrrr}
    
    \hline
     & & \multicolumn{3}{l}{Male} 
    & \multicolumn{3}{l}{Female} 
    & \multicolumn{3}{c}{BSc} 
    & \multicolumn{3}{c}{MSc} \\
    SrM & Num && $\mu$  &$\sigma$ && Num & $\mu$ & $\sigma$ & Num & $\mu$ & $\sigma$ & Num & $\mu$ & $\sigma$ \\
    
    \hline
    Competent & 37 && 3.0 & 1.0 && 11 & 2.7 & 1.0 & 27 & 3.0 & 1.0 & 21 & 3.0 & 1.0 \\
    Skillful & 37 && 3.0 & 1.0 && 11 & 3.0 & 1.0 &27 & 3.1 & 1.0 & 21 & 3.0 & 1.0 \\
    Knowledgeable & 37 && 3.1 & 1.0 && 11 & 3.3 & 1.0 & 27 & 3.1 & 1.0 & 21 & 3.0 & 1.0\\
    Moral & 37 && 2.6 & 1.0 && 11 & 3.1 & 1.4 & 27 & 3.0 & 1.1 & 21 & 3.0 & 1.0\\
    Trustworthy & 37 && 2.6 & 1.0 && 11 & 3.1 & 1.0 & 27 & 3.0 & 1.0 & 21 &3.0 & 1.0 \\
    \hline
    
     & & \multicolumn{3}{l}{Male} 
    & \multicolumn{3}{l}{Female} 
    & \multicolumn{3}{c}{BSc} 
    & \multicolumn{3}{c}{MSc}\\
    
    SrF & Num && $\mu$  &$\sigma$ && Num & $\mu$ & $\sigma$ & Num & $\mu$ & $\sigma$ & Num & $\mu$ & $\sigma$\\
    
    \hline
    Competent & 36 && 3.0 & 1.0 && 11 & 3.1 & 1.0 &27 & 3.0 & 1.0 & 20 & 3.1 & 1.0\\
    Skillful & 36 && 3.0 & 1.0 && 11 & 3.1 & 1.0 & 27 & 3.1 & 1.0 & 20 & 3.0 & 1.0 \\
    Knowledgeable & 36 && 3.1 & 1.0 && 11 & 2.9 & 1.0 & 27 & 3.0 & 1.0 & 20 & 3.2 & 1.0\\
    Moral & 36 && 3.0 & 1.2 && 11 & 3.5 & 1.0 & 27 & 3.0 & 1.0 & 20 & 3.4 & 1.2 \\
    Trustworthy & 36 && 2.9 & 1.0 && 11 & 3.0 & 1.0 & 27 & 3.0 & 1.0 & 20 & 3.0 & 1.0 \\
    \hline
   
    & & \multicolumn{3}{l}{Male} 
    & \multicolumn{3}{l}{Female} 
    & \multicolumn{3}{c}{BSc} 
    & \multicolumn{3}{c}{MSc} \\
    
    JrM & Num && $\mu$  &$\sigma$ && Num & $\mu$ & $\sigma$ & Num & $\mu$ & $\sigma$ & Num & $\mu$ & $\sigma$\\
    
    \hline
    Competent & 36 && 2.9 & 1.0 && 11 & 3.3 & 1.0 & 20 & 3.0 & 1.0 & 27 & 3.0 & 1.0  \\
    Skillful & 36 && 3.0 & 1.0 && 11 & 3.0 & 1.0 & 20 & 3.0 & 1.0 & 27 &3.0 & 1.0  \\
    Knowledgeable & 36 && 3.0 & 1.0 && 11 & 3.2 & 1.0 & 20 &3.2 & 1.0 & 27 & 3.0 & 1.0 \\
    Moral & 36 && 3.0 & 1.1 && 11 & 3.2 & 1.0 & 20 &3.0 & 1.0 & 27 & 3.0 & 1.0 \\
    Trustworthy & 36 && 3.0 & 1.0 && 11 & 3.0 & 0.4 & 20 & 3.0 & 1.0& 27 & 3.0 & 1.0 \\
    \hline
    
     & & \multicolumn{3}{l}{Male} 
    & \multicolumn{3}{l}{Female} 
    & \multicolumn{3}{c}{BSc} 
    & \multicolumn{3}{c}{MSc} \\
    
    JrF & Num && $\mu$  &$\sigma$ && Num & $\mu$ & $\sigma$ & Num & $\mu$ & $\sigma$ & Num & $\mu$ & $\sigma$ \\
    
    \hline
    Competent & 35 && 3.1 & 1.0 && 11 & 3.0 & 1.0 & 25 & 3.0 & 1.0 & 21 & 3.2 & 1.0 \\
    Skillful & 35 && 3.2 & 1.0 && 11 & 3.0 & 1.1 & 25 & 3.0 & 1.0& 21 &3.2 & 1.0 \\
    Knowledgeable & 35 && 3.2 & 1.1 && 11 & 3.0 & 1.1 & 25 & 3.1 & 1.0 & 21 & 3.2 & 1.2\\
    Moral & 35 && 3.0 & 1.2 && 11 &  3.0 & 1.0 & 25 & 3.0 & 1.2 & 21 & 3.0 & 1.0 \\
    Trustworthy & 35 && 2.5 & 1.0 && 11 & 3.0 & 1.0 & 25 & 2.3 & 1.0 & 21 &  3.0 & 1.0 \\
    \hline
    
    \end{tabular}
    \label{tab:analyst}
\end{table*}


\begin{table*}[h]
    \caption{Perception of Outcomes Based on Gender (Male, Female) and Level of education (BSc, MSc) of Participants (\textit{SrM- senior male, SrF- senior female, JrM- junior male, JrF- junior female}, 5-point likert scale).}
    \centering
    \footnotesize
    \label{tab:recognized:threats}
    \begin{tabular}{lrrrrrrrr|rrrrrr}
    
    \hline
     & & \multicolumn{3}{l}{Male} 
    & \multicolumn{3}{l}{Female} 
    & \multicolumn{3}{c}{BSc} 
    & \multicolumn{3}{c}{MSc}\\
    SrM & Num && $\mu$  &$\sigma$ && Num & $\mu$ & $\sigma$ & Num & $\mu$ & $\sigma$ & Num & $\mu$ & $\sigma$ \\
    
    \hline
 Ethical & 37 && 2.4 & 1.2 && 11 & 3.1 & 1.1 & 27 & 3.0 & 1.3 & 21 & 2.5 & 1.1\\
 Reliable & 37 && 2.5 & 1.1 && 11 & 3.1 & 1.0 & 27 & 3.0 & 1.1 & 21 & 2.4 & 1.0\\
 Responsible& 37 && 2.6 & 1.1 && 11 & 2.7 & 1.0 & 27 & 3.0 & 1.1 & 21 & 2.5 & 1.0\\
 Trustworthy& 37 && 2.3 & 1.0 && 11 & 2.7 & 1.0 & 27 &3.0 & 1.1& 21 & 2.2 & 1.1\\
 Overall acceptability & 37 && 2.8 & 1.1 && 11 & 2.7 & 1.1 & 27 & 3.1 & 1.1 & 21 & 2.4 & 1.0\\
    \hline
     & & \multicolumn{3}{l}{Male} 
    & \multicolumn{3}{l}{Female} 
    & \multicolumn{3}{c}{BSc} 
    & \multicolumn{3}{c}{MSc} \\
    SrF & Num && $\mu$  &$\sigma$ && Num & $\mu$ & $\sigma$ & Num & $\mu$ & $\sigma$ & Num & $\mu$ & $\sigma$ \\
    
    \hline
 Ethical & 36 && 3.1 & 1.3 &&  11 & 3.0 & 1.3 & 27 & 3.0 & 1.2 & 20 & 3.3 & 1.4\\
 Reliable & 36 && 2.5 & 1.0 && 11 & 2.5 & 1.0 & 27 & 3.0 & 1.0 & 20 & 2.4 & 1.0\\
 Responsible& 36 && 2.8 & 1.3 && 11 & 2.9 & 1.2 & 27 & 3.0 & 1.2 & 20 & 3.1 & 1.2\\
 Trustworthy& 36 && 2.5 & 1.1 && 11 & 2.6 & 1.0 & 27 & 2.4 & 1.0 & 20 & 3.0 & 1.1\\
 Overall acceptability & 36 && 2.8 & 1.3 &&11 & 3.1 & 1.0 & 27 & 3.0 & 1.2 & 20 & 3.1 & 1.2\\
    \hline
    
     & & \multicolumn{3}{l}{Male} 
    & \multicolumn{3}{l}{Female} 
    & \multicolumn{3}{c}{BSc} 
    & \multicolumn{3}{c}{MSc}\\
    JrM & Num && $\mu$  &$\sigma$ && Num & $\mu$ & $\sigma$ & Num & $\mu$ & $\sigma$ & Num & $\mu$ & $\sigma$  \\
    
    \hline
 Ethical & 36 && 2.8 & 1.1 && 11 & 2.8 & 1.0 & 20 & 3.0 & 1.0 & 27 & 3.0 & 1.1 \\
 Reliable & 36 && 2.6 & 1.0 && 11 & 2.6 & 1.1 & 20 & 3.0 & 1.0  & 27 & 2.4 & 1.0\\
 Responsible& 36 && 2.7 & 1.2 && 11 & 3.0 & 1.0 & 20 & 3.0 & 1.3 & 27 & 3.0 & 1.0  \\
 Trustworthy& 36 && 2.4 & 1.0 && 11 & 3.0 & 1.0 & 20 & 3.0 & 1.0  & 27 & 2.4 & 1.0 \\
 Overall acceptability & 36 && 2.7 & 1.0 && 11 & 3.0 & 1.0 & 20 & 3.0 & 1.0 & 27 & 3.0 & 1.0 \\
    \hline
    
    & & \multicolumn{3}{l}{Male} 
    & \multicolumn{3}{l}{Female} 
    & \multicolumn{3}{c}{BSc} 
    & \multicolumn{3}{c}{MSc}\\
    JrF & Num && $\mu$  &$\sigma$ && Num & $\mu$ & $\sigma$ & Num & $\mu$ & $\sigma$ & Num & $\mu$ & $\sigma$ \\
    
    \hline
 Ethical & 35 && 2.8 & 1.4 && 11 & 2.8 & 1.3 & 25 & 3.0 & 1.4 & 21 & 3.0& 1.3 \\
 Reliable & 35 && 3.0 & 1.1 && 11 & 2.6 & 1.0 & 25 & 3.0 & 1.0 & 21 & 3.0 & 1.0 \\
 Responsible& 35 && 2.7 & 1.2 && 11 & 2.5 & 1.0 & 25 & 3.0 & 1.1 & 21 & 3.0 & 1.0 \\
 Trustworthy& 35 && 2.5 & 1.0 && 11 & 2.5 & 1.1 & 25 & 2.3 & 1.0 & 21 & 3.0 & 1.0 \\
 Overall acceptability & 35 && 2.9 & 1.0 && 11 & 3.0 & 1.0 & 25 & 3.0 & 1.2 & 21 & 3.0 & 1.0\\
    \hline
    
\end{tabular}
\label{tab:outcomes}
\end{table*}



\subsection{Detailed observations}
\label{subsec:analyst}
\paragraph{\normalfont{\textbf{Perception of analyst persona}}}For the perception of analyst persona, participants were asked to rate (on a Likert scale with 5 levels) how competent, skillful, knowledgeable, moral, and trustworthy they found the analyst persona.

Table \ref{tab:analyst} is a summary of the mean and standard deviation for the perception of each analyst persona based on the gender and education level of the participants. 
From the left column, we observed that female participants were on average neutral when rating their perception of each vignette persona. On the other hand, male students appeared to disagree for what concerns the morality (for SrM) and trustworthiness (for SrM ad JrF). For what concerns the education level of the participants, both bachelor and master students rated the randomly assigned vignettes more neutral for each measured variables. 

\paragraph{\normalfont{\textbf{Perception of security mitigations}}} To check for participants perception of the security mitigations presented to them, they were required to rate their confidence of the security solution presented to them by the analyst personas. To this end, participants were asked to rate (on a Likert scale with 5 levels) how ethical, reliable, responsible, trustworthy, and overall acceptable they found the proposed security mitigation.

%

Table \ref{tab:outcomes} is a summary of the responses which were reviewed with regard to the gender and educational background of the participants. The left column shows some differences between the responses of male and female participants. We observed that for what concerns ethical and the overall acceptability of the outcomes, both male and female participants were more neutral than the other measures. For the differences with regard to the level of education, we observed that Bachelor students were neutral in their responses, to the outcomes proposed by all four vignette personas. On the other hand, master students were neutral except for what concerns SrM (Frank, senior analyst) where they appeared to be in disagreement.

However, the observed differences in Tables \ref{tab:analyst} and \ref{tab:outcomes} are small and may not infer any significant equivalence or differences. In order to check for statistical significance, we aggregated the treated measures into a scale using Chronbach alpha, see Section \ref{sec:results}.

\begin{table}[h]
\caption{Test of equivalence of participants' perception based on their gender and education level}\label{tab:scale_TOST}
\centering
\resizebox{0.49\textwidth}{!}{
\begin{tabular}{ll|rr|rr}
\toprule
& & \multicolumn{2}{c}{Analyst}  & \multicolumn{2}{c}{Mitigation}  \\

 &  $\alpha =0.05$ & $MWU$ & $P$ & $MWU$ & $P$ \\
\midrule
  Participants' gender  & {$Female$ - $\delta$} $<$ {$Male$}   & 1119.0 & \textbf{3.8e-11*} & 1933.5 & \textbf{4.5e-05*} ($p_1$)\\

  & {$Male$} $<$ {$Female$ + $\delta$} & 960.0  & \textbf{1.1e-12} & 1283.0 &  \textbf{1.1e-09}  ($p_2$)\\

\hline
        
   Level of education &  {$BSc$ - $\delta$} $<$ {$MSc$} & 1463.0 & 1.2e-15  &  2317.5 & \textbf{1.0e-08}  ($p_1$) \\
    & {$MSc$} $<$ {$BSc$ + $\delta$} & 1674.0 &  \textbf{1.0e-13*}  &  2328.5 & \textbf{1.1e-08*}  ($p_2$)\\  

    \bottomrule
\end{tabular}}
\end{table}

\begin{table}[h]
\caption{Test of equivalence on participants' perception of the case study}\label{tab:maxx_TOST}
\centering
\resizebox{0.49\textwidth}{!}{
\begin{tabular}{lc|rr}
\toprule
& & \multicolumn{2}{c}{MaxxUpload} \\

 &  $\alpha =0.05$ & $MWU$ & $P$ \\
\midrule
 Participants' gender  & {$Female$ - $\delta$} $<$ {$Male$}  & 1642.0 & \textbf{4.6e-07}  ($p_1$)\\

  & {$Male$} $<$ {$Female$ + $\delta$} & 2950.0 & 0.24 ($p_2$)\\

\hline
        
Level of education    &  {$BSc$ - $\delta$} $<$ {$MSc$} & 3673.0 & \textbf{0.02*} ($p_1$) \\
    & {$MSc$} $<$ {$BSc$ + $\delta$}  & 2869.0 &  \textbf{1.5e-05} ($p_2$)\\  

\hline
    Mitigation  &  {\textit{Traffic blocking} - $\delta$} $<$ {$Malware$}   & 3028.0   & \textbf{8.4e-05}  ($p_1$) \\
    & {$Malware$} $<$ {\textit{Traffic blocking} + $\delta$}  & 3529.0  & \textbf{0.008*} ($p_2$)\\ 
    \bottomrule
\end{tabular}}
\end{table}

\begin{table}[h]
\caption{Test of equivalence on perception based received mitigation}\label{tab:TOST_mitigation}
\centering
\resizebox{0.49\textwidth}{!}{
\begin{tabular}{c|rr|rr}
\toprule
& \multicolumn{2}{c}{Analyst}  & \multicolumn{2}{c}{Mitigation}  \\

 $\alpha =0.05$ & $MWU$ & $P$ & $MWU$ & $P$ \\
\midrule
 {\textit{Traffic blocking} - $\delta$} $<$ {$Malware$}  & 1567.0  & \textbf{9.6e-15}  & 3929.5 & 0.09  ($p_1$) \\
{$Malware$} $<$ {\textit{Traffic blocking} + $\delta$}  & 1616.0 & \textbf{2.6e-14*} & 989.5 &  \textbf{1.7e-20} ($p_2$)\\ 
    \bottomrule
\end{tabular}}
\end{table}

\end{document}